\documentclass[9pt,twocolumn]{article}

\usepackage[margin=0.7in]{geometry}
\usepackage{textcomp}
\usepackage{microtype}
\usepackage{graphicx}
\usepackage{subcaption}
\usepackage{booktabs}
\usepackage{colortbl}
\usepackage{xcolor}
\usepackage{hyperref}
\usepackage{url}

\usepackage{amsmath}
\usepackage{amssymb}
\usepackage{mathtools}
\usepackage{amsthm}

\usepackage{algorithm}
\usepackage{algorithmic}
\usepackage{authblk}
\usepackage[capitalize,noabbrev]{cleveref}

\definecolor{ourscolor}{gray}{0.92}

\usepackage{float}
\usepackage{stfloats}

\theoremstyle{plain}

\theoremstyle{definition}

\theoremstyle{remark}

\usepackage{mathpazo}  
\usepackage[T1]{fontenc}

\title{Soft Clustering Anchors for Self-Supervised Speech Representation Learning in Joint Embedding Prediction Architectures}

\author{
  \textbf{Georgios Ioannides}$^{1,3,7}$ \quad
  \textbf{Adrian Kieback}$^{3,*}$ \quad
  \textbf{Judah Goldfeder}$^{4,*}$ \quad
  \textbf{Linsey Pang}$^{5}$\\
  \textbf{Aman Chadha}$^{3,6}$ \quad
  \textbf{Aaron Elkins}$^{3}$ \quad
  \textbf{Yann LeCun}$^{2}$ \quad
  \textbf{Ravid Shwartz-Ziv}$^{2}$\vspace{1em}\\
  {\small
  $^{1}$Carnegie Mellon University \quad
  $^{2}$New York University \quad
  $^{3}$James Silberrad Brown Center for AI\\
  $^{4}$Columbia University \quad
  $^{5}$Northeastern University \quad
  $^{6}$Stanford University \quad
  $^{7}$Amazon GenAI$^{\dagger}$ \vspace{0.5em}\\
  $^{*}$Equal contribution \quad Correspondence: \texttt{gioannid@alumni.cmu.edu}
  }
}

\date{}

\begin{document}

\maketitle

\renewcommand{\thefootnote}{\fnsymbol{footnote}}
\footnotetext[2]{Work does not relate to position at Amazon.}

\begin{abstract}
Joint Embedding Predictive Architectures (JEPA) offer a promising approach to self-supervised speech representation learning, but suffer from representation collapse without explicit grounding. We propose GMM-Anchored JEPA, which fits a Gaussian Mixture Model once on log-mel spectrograms and uses its frozen soft posteriors as auxiliary targets throughout training. A decaying supervision schedule allows GMM regularization to dominate early training before gradually yielding to the JEPA objective. Unlike HuBERT and WavLM, which require iterative re-clustering, our approach clusters input features once with soft rather than hard assignments. On ${\sim}$50k hours of speech, GMM anchoring improves ASR (28.68\% vs.\ 33.22\% WER), emotion recognition (67.76\% vs.\ 65.46\%), and slot filling (64.7\% vs.\ 59.1\% F1) compared to a WavLM-style baseline with matched compute. Cluster analysis shows GMM-anchored representations achieve up to 98\% entropy compared to 31\% for WavLM-style, indicating substantially more uniform cluster utilization. Code is made available\footnote{ \url{https://github.com/gioannides/clustering-anchored-jepa}}.
\end{abstract}

\textbf{Keywords:} Self-supervised learning, speech representation, JEPA, Gaussian mixture models

\section{Introduction}

Self-supervised learning has become the dominant paradigm for speech representation learning \cite{baevski2020wav2vec, hsu2021hubert, chen2022wavlm}. Joint Embedding Predictive Architectures (JEPA) \cite{lecun2022jepa, assran2023ijepa} offer a promising direction by predicting latent representations rather than reconstructing inputs or contrasting samples, avoiding the need for iterative re-clustering that dominates current speech SSL methods. However, when applied to speech, we find that JEPA suffers from severe representation collapse: without precautions, JEPAs map inputs to degenerate representations \cite{drozdov2024videojepa, assran2023ijepa}. In our experiments, unanchored JEPA produces near-random predictions on downstream ASR, indicating that the encoder discards phonetic distinctions entirely. Our goal is not to achieve state-of-the-art absolute performance, but to demonstrate that frozen acoustic anchors can stabilize JEPA training for speech, eliminating the need for iterative re-clustering while maintaining competitive downstream performance.

Existing solutions to representation collapse in speech SSL rely on iterative offline clustering. HuBERT \cite{hsu2021hubert} and WavLM \cite{chen2022wavlm} fit k-means on intermediate representations, retrain the model with new targets, and repeat. This costly pipeline multiplies training compute. These methods also use hard cluster assignments, discarding uncertainty at acoustic boundaries.

\begin{figure*}[t]
\centering
\includegraphics[width=0.8\textwidth]{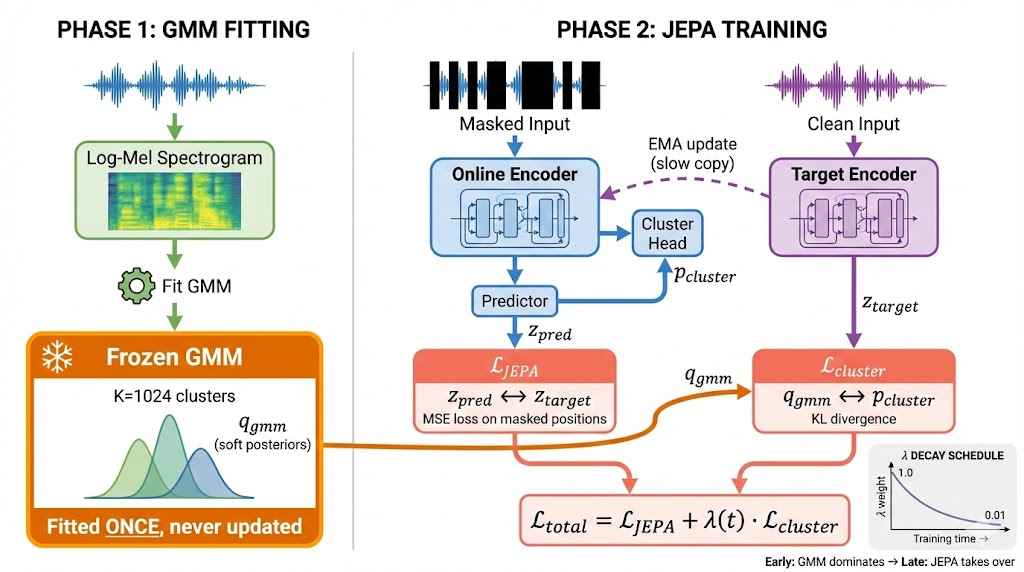}
\caption{\textbf{GMM-Anchored JEPA: one-time clustering replaces iterative re-training.} \textbf{Phase 1:} A GMM is fitted once on log-mel features. \textbf{Phase 2:} The encoder trains with two objectives: predicting masked EMA teacher latents (JEPA loss) and matching frozen GMM posteriors (cluster loss). $\lambda(t)$ decays from 1.0 to 0.01.}
\label{fig:overview}
\end{figure*}

We propose to anchor JEPA training with a frozen Gaussian Mixture Model fitted once on log-mel spectrograms (Figure~\ref{fig:overview}). Because the GMM is frozen, it provides stable acoustic targets that cannot co-adapt with the encoder. An auxiliary Kullback-Leibler (KL) divergence loss matches a cluster head's output to the GMM's soft posteriors, with weight decaying from $\lambda=1.0$ to $\lambda=0.01$ over training so that cluster supervision dominates early while the JEPA objective gradually takes over. Unlike HuBERT and WavLM, which require iterative re-clustering of intermediate representations, our GMM is fitted once on input features. And unlike hard k-means assignments, GMM posteriors preserve uncertainty at cluster boundaries, which we hypothesize provides richer supervision signal.

On approximately 50k hours of speech, GMM-anchored JEPA reduces WER by 14\% relative on LibriSpeech \cite{panayotov2015librispeech} compared to a WavLM-style baseline with matched compute, while also improving emotion recognition and slot filling. Cluster analysis reveals that GMM-anchored representations achieve up to 98\% entropy versus 31\% for WavLM-style, indicating substantially more uniform cluster utilization.

\paragraph{Contributions.}
\begin{itemize}
    \item We identify representation collapse in JEPA-based speech SSL and propose \emph{frozen GMM anchoring} with a decaying supervision schedule, replacing iterative re-clustering with one-time soft clustering of input features.
    \item We demonstrate that residual GMM supervision ($\lambda=0.01$) provides ongoing regularization throughout training: removing it causes representations to collapse even after initial grounding, showing that GMM anchoring is not merely initialization but continuous stabilization.
    \item We show that GMM anchoring generalizes across encoder architectures (Conformer and Transformer), improving downstream performance on ASR, emotion recognition, and slot filling while achieving up to 98\% cluster entropy compared to 31\% for WavLM-style baselines.
\end{itemize}

\section{Related Work}

\paragraph{Self-Supervised Speech Learning.}
Contrastive methods (CPC \cite{oord2018cpc}, wav2vec 2.0 \cite{baevski2020wav2vec}) distinguish positive from negative samples, but require negative sampling strategies and codebook diversity losses to avoid collapse. Masked prediction methods (HuBERT \cite{hsu2021hubert}, WavLM \cite{chen2022wavlm}) use offline k-means clustering for discrete targets, iteratively re-clustering intermediate layers, which multiplies training compute and requires careful scheduling. BEST-RQ \cite{chiu2022bestrq} avoids re-clustering by using a frozen random-projection quantizer, but provides random rather than acoustically informed targets. JEPA \cite{lecun2022jepa, assran2023ijepa} predicts latent representations directly without reconstruction or negatives, but suffers from representation collapse in speech without external grounding.

\paragraph{Self-Distillation and Collapse.}
BYOL \cite{grill2020byol} and DINO \cite{caron2021dino} showed that self-distillation with an EMA teacher can learn without negatives, using centering, sharpening, or asymmetric architectures to prevent collapse. data2vec \cite{baevski2022data2vec} applies this to speech, predicting averaged teacher representations across layers. However, representation collapse remains a challenge in self-supervised learning \cite{jing2022understanding}, particularly for speech where the EMA teacher provides no spectral grounding. The issue is that EMA-based teachers in speech SSL provide only temporal smoothing without acoustic anchoring: the teacher's representations drift alongside the student, offering no stable reference to ground phonetic distinctions. Our work addresses this by anchoring JEPA with frozen GMM supervision that cannot co-adapt with the encoder.

\paragraph{Clustering Supervision.}
HuBERT and WavLM use hard k-means assignments that discard uncertainty at cluster boundaries \cite{macqueen1967kmeans, bezdek1981fuzzy}. When a frame lies between two phonetic categories, hard assignment forces an arbitrary choice, losing information about the acoustic ambiguity. GMMs have long been used in speech processing \cite{bishop2006pattern}, where soft posteriors naturally model the gradual transitions between phonemes. Like BEST-RQ, we use a frozen clustering model to avoid iterative re-training, but unlike BEST-RQ's random projections, our GMM provides acoustically meaningful soft posteriors fitted to the input feature space. The frozen GMM serves as an external anchor that grounds representations in spectral structure while the decaying supervision schedule allows the JEPA objective to refine higher-level abstractions.

\section{Method}

\subsection{Framework Overview}

Our framework has two phases. In the first, a Gaussian Mixture Model is fitted once on log-mel spectrograms, producing soft posterior targets over $K$ clusters. In the second, a student encoder $f_\phi$ trains jointly with two objectives: (1) predicting masked latent representations from an EMA teacher $f_{\phi'}$, and (2) matching a cluster head's output to the frozen GMM posteriors. The GMM is never updated during training.

\subsection{Phase 1: GMM Fitting}

A $K$-component diagonal-covariance GMM is fitted on log-mel features $M = \log(\text{MelSpec}(\mathbf{x}) + \epsilon)$. We use log-mel spectrograms rather than raw waveforms or learned features because they provide a stable, low-dimensional acoustic representation that does not change during training. Diagonal covariance keeps fitting tractable on large corpora while still capturing per-frequency variance. The GMM produces soft posterior assignments:
\begin{equation}
q_k(\mathbf{m}) = \frac{\pi_k \mathcal{N}(\mathbf{m}; \boldsymbol{\mu}_k, \boldsymbol{\sigma}_k^2)}{\sum_{j} \pi_j \mathcal{N}(\mathbf{m}; \boldsymbol{\mu}_j, \boldsymbol{\sigma}_j^2)}.
\end{equation}

Unlike hard k-means labels, these posteriors assign nonzero probability to multiple clusters, so frames near a boundary are not forced into a single partition.

\subsection{Phase 2: Joint Training}

\paragraph{JEPA Loss.} A predictor $h_\psi$ predicts EMA teacher latents at masked positions:
\begin{equation}
\mathcal{L}_{\text{JEPA}} = \frac{1}{|\mathcal{M}|} \sum_{t \in \mathcal{M}} \| h_\psi(\tilde{z}_{\text{student}})_t - z_{\text{teacher},t} \|^2.
\end{equation}

\paragraph{Cluster Loss.} A cluster head maps encoder outputs to $K$ logits, trained to match GMM posteriors:
\begin{equation}
\mathcal{L}_{\text{cluster}} = \frac{1}{|\mathcal{M}|} \sum_{t \in \mathcal{M}} \text{KL}(q_{\text{gmm},t} \| p_{\text{cluster},t}).
\end{equation}

\paragraph{Total Loss with Decay.}
\begin{equation}
\mathcal{L}_{\text{total}} = \mathcal{L}_{\text{JEPA}} + \lambda(t) \cdot \mathcal{L}_{\text{cluster}},
\end{equation}
where $\lambda(t)$ decays linearly from 1.0 to 0.01. Early training is thus dominated by GMM supervision, grounding representations in acoustic structure. As $\lambda$ decays, the JEPA objective takes over while residual anchoring prevents drift. We use KL divergence because both $q_{\text{gmm}}$ and $p_{\text{cluster}}$ are distributions over $K$ clusters, and KL penalizes mismatched modes more heavily than L2. The residual weight $\lambda_{\text{end}}=0.01$ is retained rather than reduced to zero: as shown in Section~\ref{sec:analysis}, removing the anchor entirely causes representation collapse.

\subsection{Denoising Augmentation}

Denoising augmentation with energy-based mixing is applied to improve the robustness of the student encoder (i.e.\ the Online Encoder being trained via gradient based optimization). The target encoder (i.e.\ the network which is a delayed copy of the Online Encoder) sees clean audio while the online encoder sees augmented audio, encouraging denoising in learned representations.

\paragraph{Noise Addition.} Given clean waveform $\mathbf{x}_{\text{clean}}$ and noise $\mathbf{n}$, the target mixing Signal-to-Noise Ratio (SNR) is set at:
\begin{equation}
\mathbf{x}_{\text{aug}} = \mathbf{x}_{\text{clean}} + \alpha \cdot \mathbf{n}, \quad \alpha = \sqrt{\frac{E_{\text{clean}}}{10^{\text{SNR}/10} \cdot E_{\text{noise}}}},
\end{equation}
where $E = \frac{1}{N}\sum_i x_i^2$ denotes energy and SNR is sampled from $[-5, 20]$ dB with probability 0.25.

\paragraph{Utterance Mixing.} A segment from another utterance is mixed with energy-based weighting:
\begin{equation}
\mathbf{x}_{\text{mix}}[t_1:t_2] = \mathbf{x}_1[t_1:t_2] + \beta \cdot \mathbf{x}_2[s_1:s_2],
\end{equation}
where $\beta = \sqrt{E_1 \cdot 10^{\rho/10} / E_2}$ with $\rho \sim [-5, 5]$ dB. Mixing probability is 0.25 with max 50\% overlap.

\begin{figure*}[h]
\centering
\begin{subfigure}[h]{0.24\textwidth}
    \includegraphics[width=\textwidth]{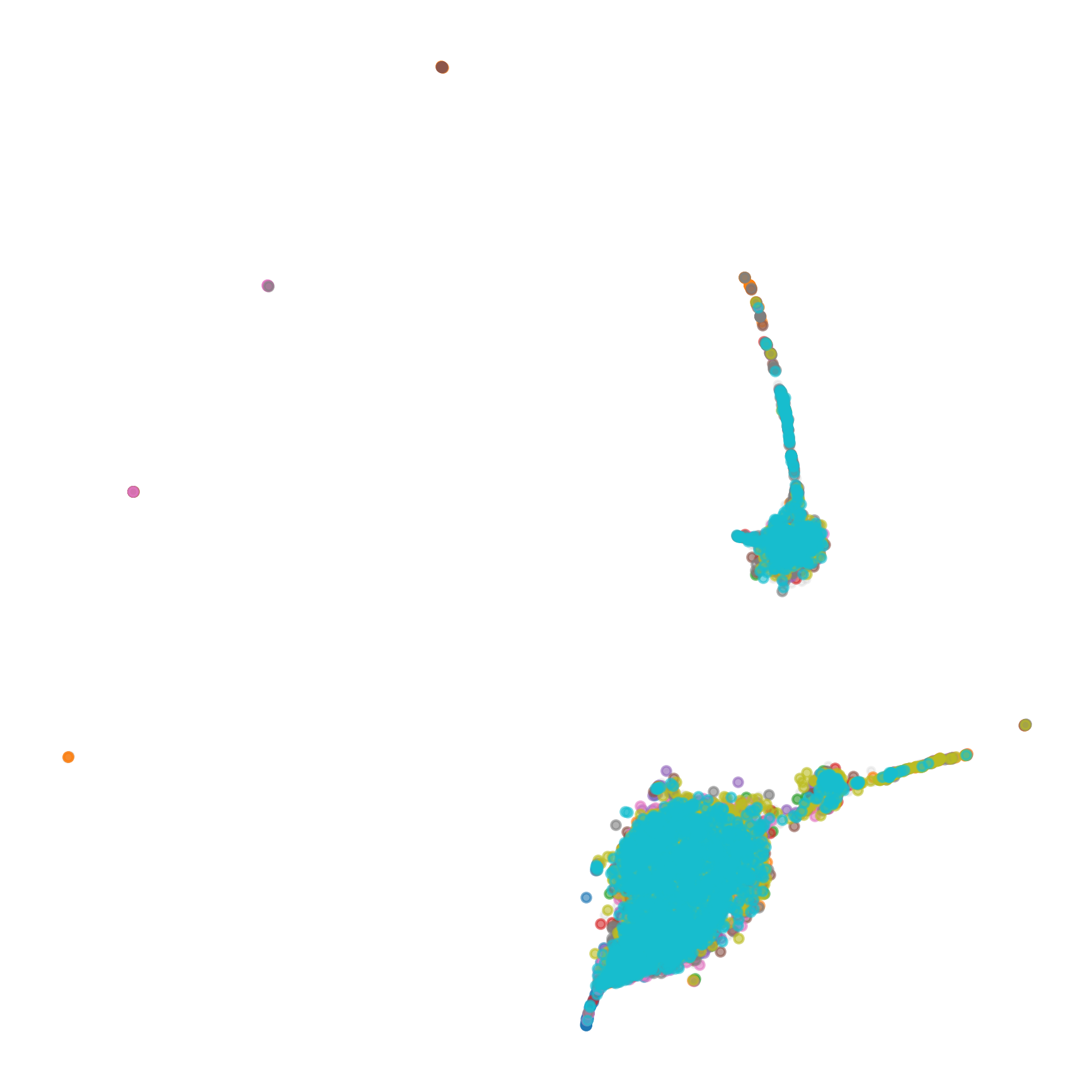}
    \caption{Pure JEPA}
\end{subfigure}
\hfill
\begin{subfigure}[h]{0.24\textwidth}
    \includegraphics[width=\textwidth]{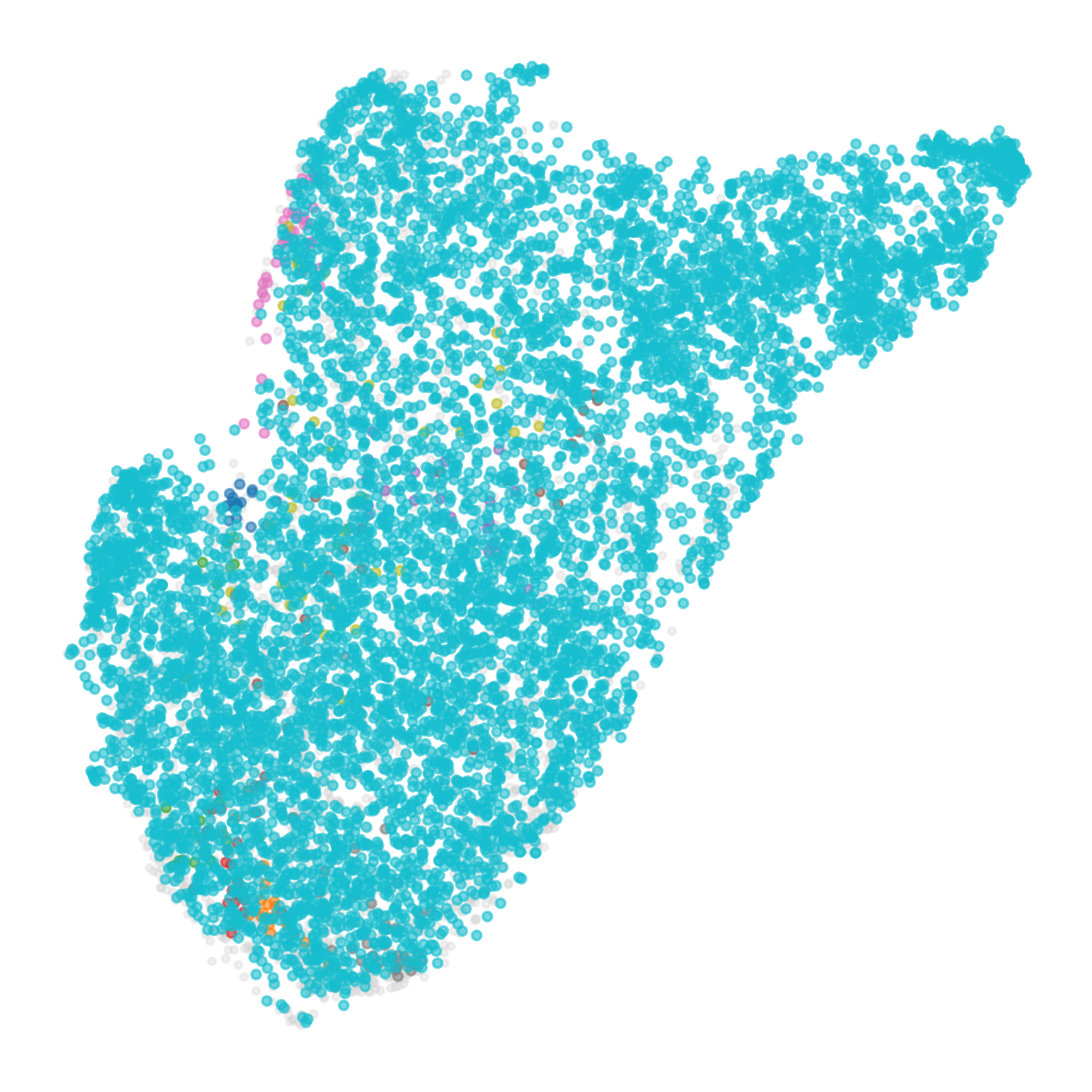}
    \caption{WavLM-style}
\end{subfigure}
\hfill
\begin{subfigure}[h]{0.24\textwidth}
    \includegraphics[width=\textwidth]{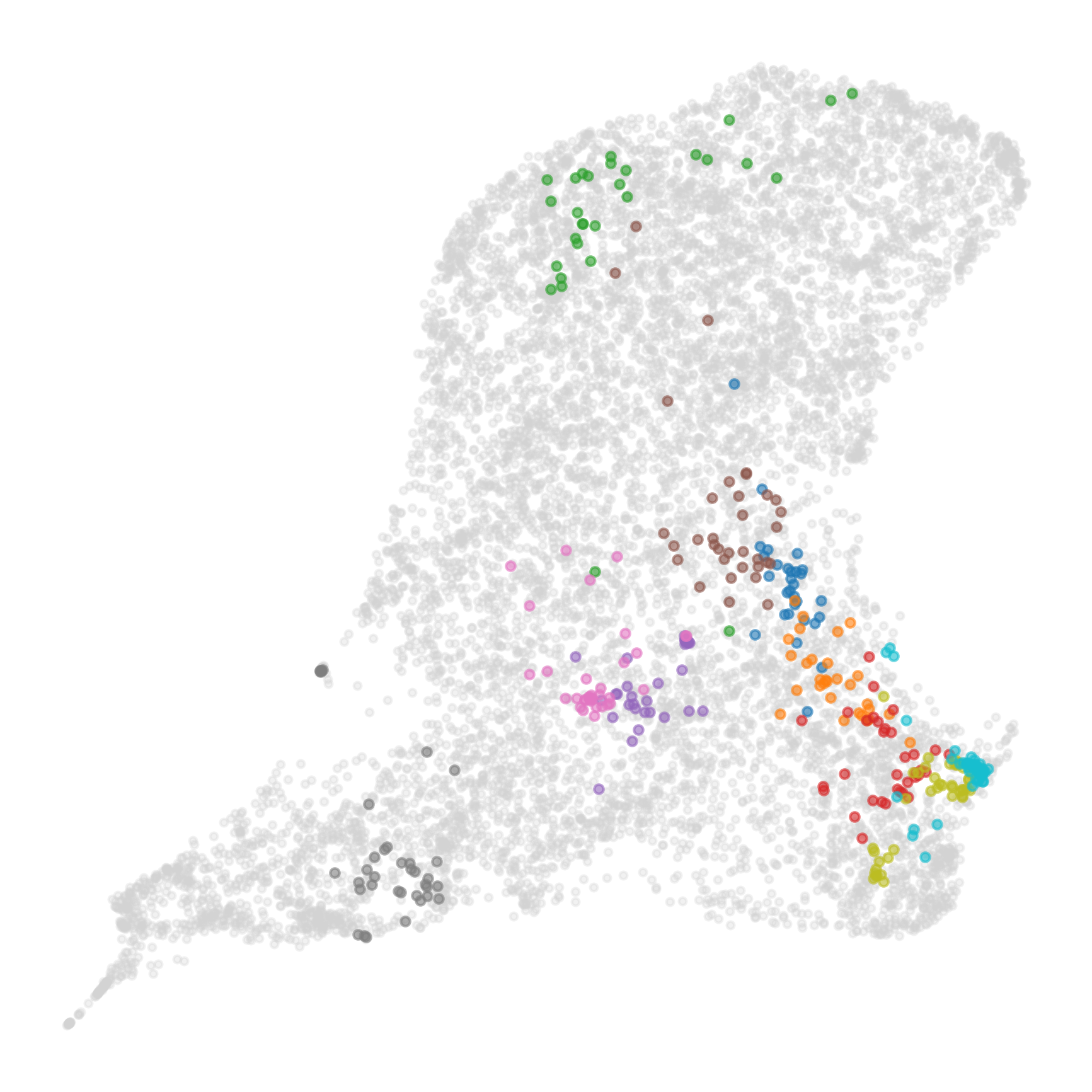}
    \caption{GMM-JEPA-T}
\end{subfigure}
\hfill
\begin{subfigure}[h]{0.24\textwidth}
    \includegraphics[width=\textwidth]{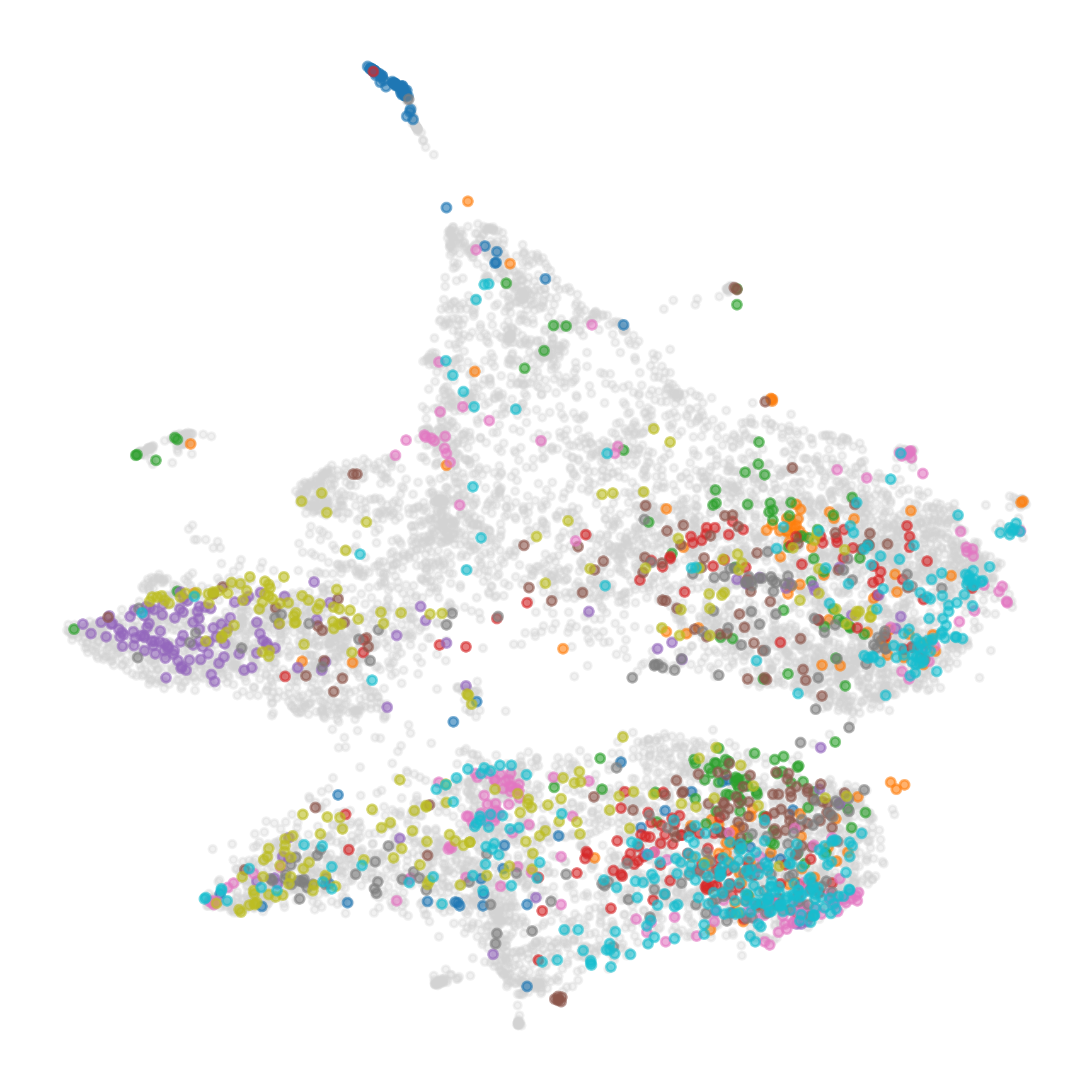}
    \caption{GMM-JEPA}
\end{subfigure}
\caption{\textbf{GMM-JEPA learns well-separated clusters; baselines collapse or overlap.} UMAP of frame-level embeddings colored by predicted cluster. \textbf{(a)} Pure JEPA collapses. \textbf{(b)} WavLM-style overlaps. \textbf{(c,d)} GMM-JEPA variants form distinct regions.}
\label{fig:umap}
\end{figure*}

\begin{figure}[h]
\centering
\includegraphics[width=0.45\textwidth]{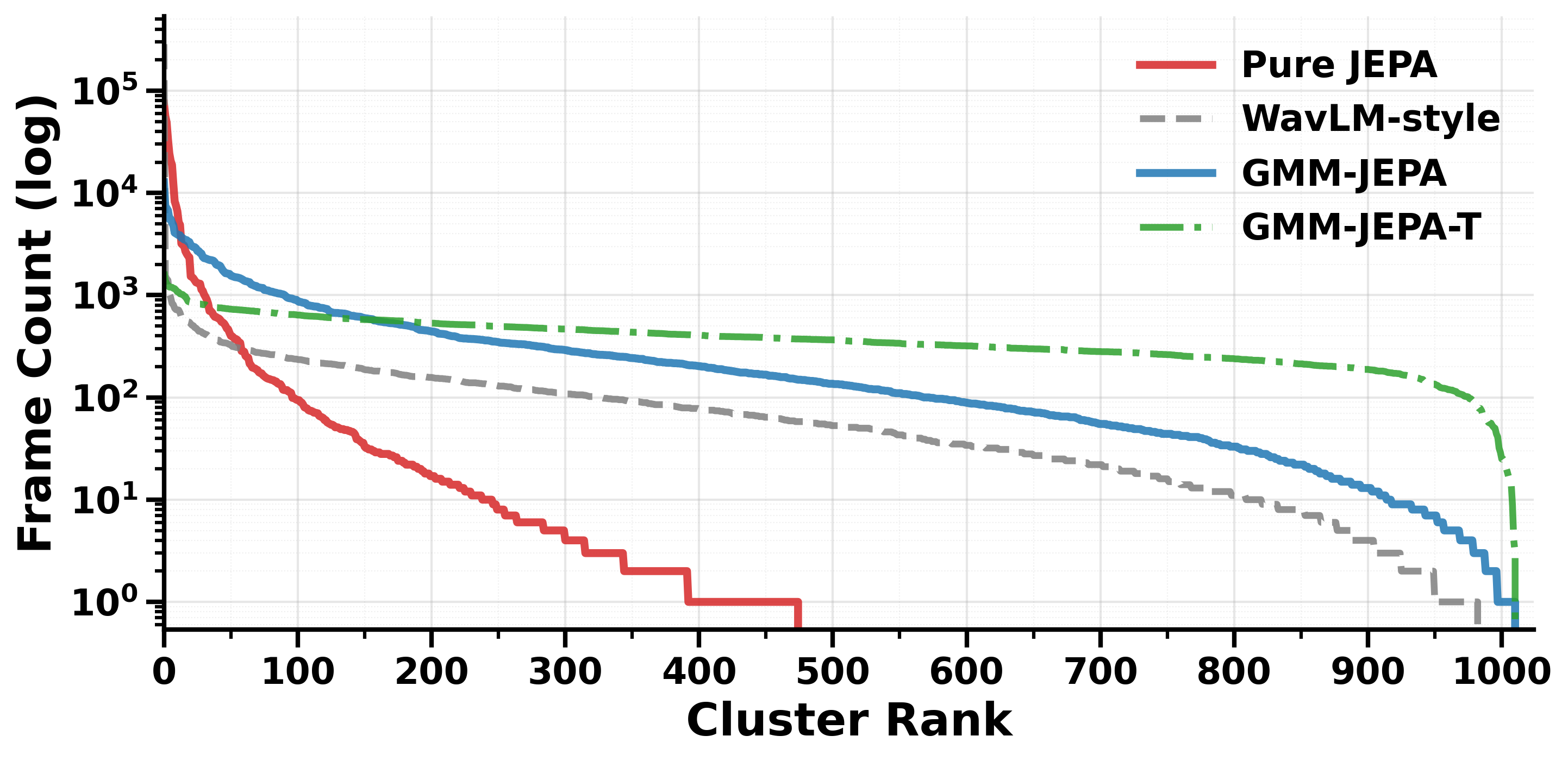}
\caption{\textbf{GMM-JEPA uses all clusters uniformly; baselines collapse to few.} Frame counts per cluster (log scale, sorted by rank). Flat = high entropy. Baselines drop steeply; GMM-JEPA stays flat.}
\label{fig:dist}
\end{figure}

\begin{figure*}[h]
\centering
\begin{subfigure}[h]{0.48\textwidth}
    \includegraphics[width=\textwidth]{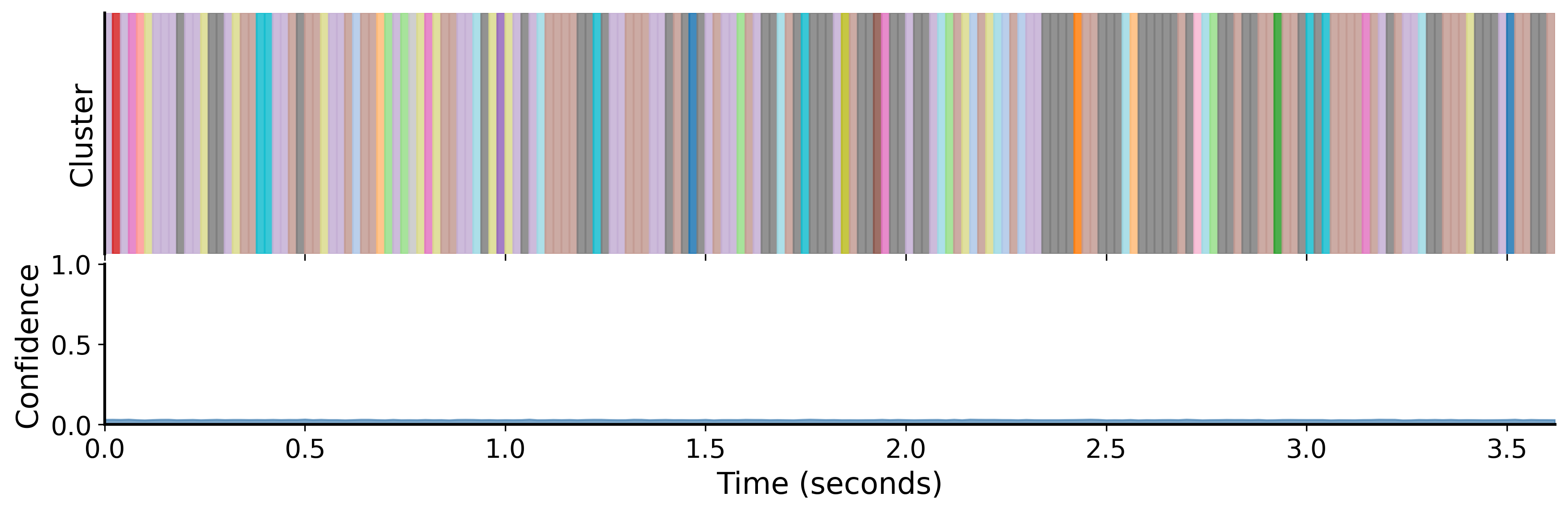}
    \caption{Pure JEPA}
\end{subfigure}
\hfill
\begin{subfigure}[h]{0.48\textwidth}
    \includegraphics[width=\textwidth]{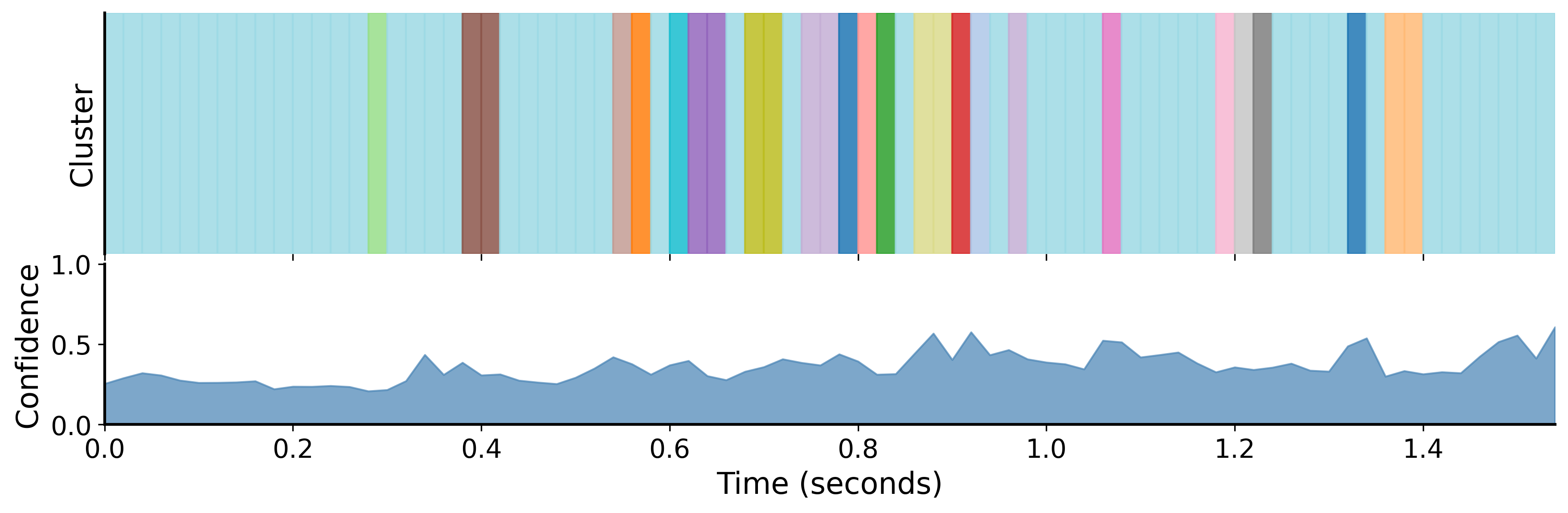}
    \caption{WavLM-style}
\end{subfigure}

\vspace{0.5em}

\begin{subfigure}[h]{0.48\textwidth}
    \includegraphics[width=\textwidth]{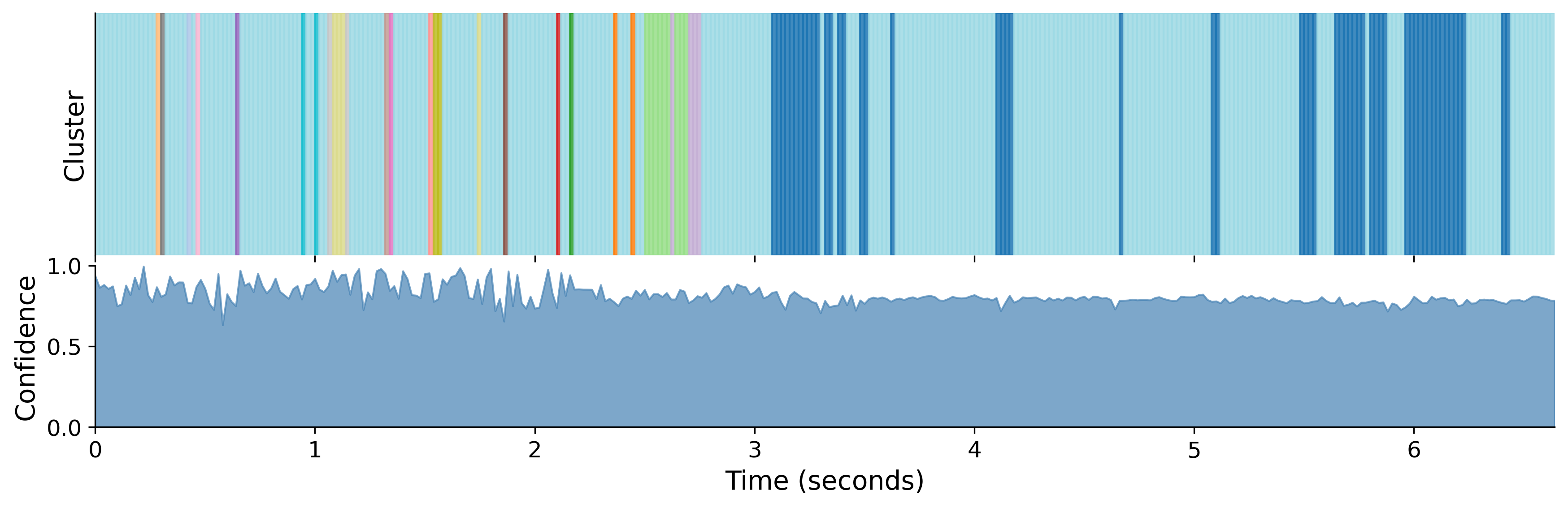}
    \caption{GMM-JEPA-T}
\end{subfigure}
\hfill
\begin{subfigure}[h]{0.48\textwidth}
    \includegraphics[width=\textwidth]{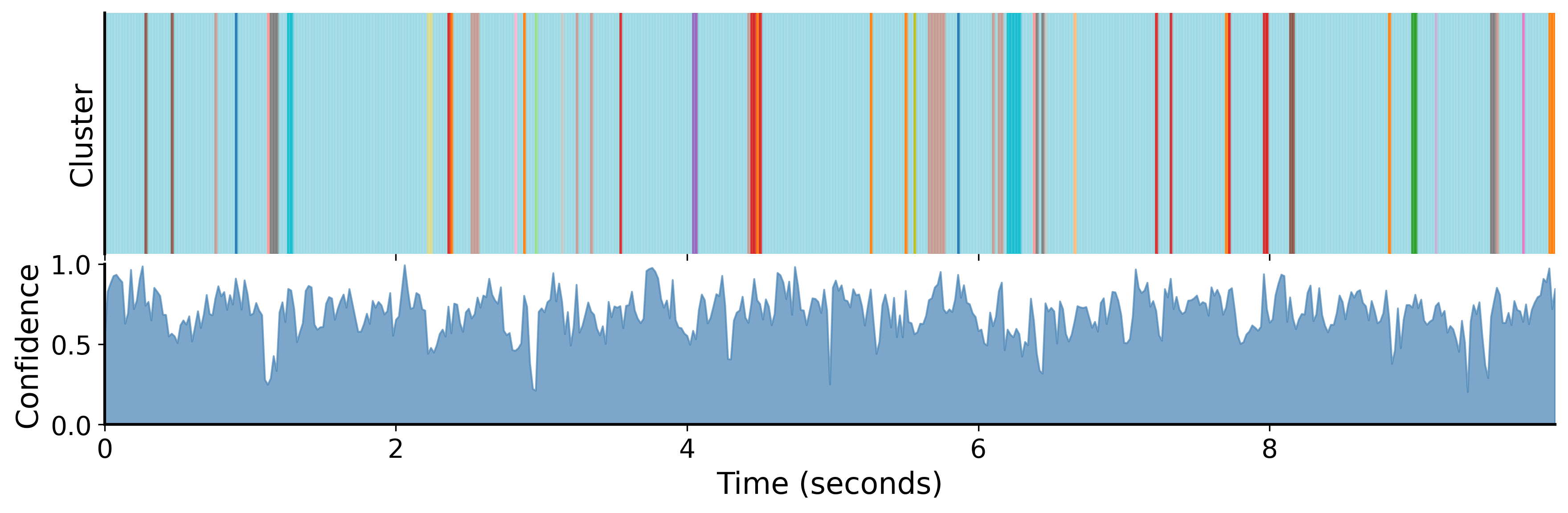}
    \caption{GMM-JEPA}
\end{subfigure}
\caption{\textbf{GMM-JEPA variants maintain stable, high-confidence clusters over time.} Top: cluster ID per frame. Bottom: confidence (1 $-$ normalized entropy). \textbf{(a)} Pure JEPA flickers rapidly with near-zero confidence, indicating degenerate representations. \textbf{(b)} WavLM-style shows moderate confidence (0.4--0.6) with frequent cluster switching. \textbf{(c)} GMM-JEPA-T maintains high confidence (0.7--0.9) with moderate cluster and sparse transitions. \textbf{(d)} GMM-JEPA also shows moderate clusters with temporally coherent spans and moderate-to-high confidence (0.5--0.8).}
\label{fig:utterance}
\end{figure*}

\subsection{Training Algorithm}

The complete training procedure is summarized in Algorithm~\ref{alg:training}. Block masking samples spans of 10-25 frames until 40-65\% masked. After training, only $f_\phi$ is retained.

\begin{algorithm}[t]
\caption{GMM-Anchored JEPA Training}
\label{alg:training}
\begin{algorithmic}[1]
\STATE \textbf{Phase 1:} Fit GMM on log-mel features
\STATE \textbf{Phase 2:}
\STATE Init: $f_\phi$, $f_{\phi'} \gets f_\phi$, $h_\psi$, cluster head, $\mathbf{t}_{\text{mask}}$
\FOR{each minibatch $\{\mathbf{x}\}$}
    \STATE $\mathbf{x}_{\text{aug}}, \mathbf{x}_{\text{clean}} \gets \text{Augment}(\mathbf{x})$
    \STATE $z_{\text{online}} \gets f_\phi(\mathbf{x}_{\text{aug}})$
    \STATE $z_{\text{target}} \gets f_{\phi'}(\mathbf{x}_{\text{clean}})$ \COMMENT{No grad}
    \STATE $q_{\text{gmm}} \gets \text{GMM.posterior}(\text{LogMel}(\mathbf{x}_{\text{clean}}))$
    \STATE $m \gets \text{SampleBlockMask}()$
    \STATE $\tilde{z} \gets m \odot z_{\text{online}} + (1-m) \odot \mathbf{t}_{\text{mask}}$
    \STATE $z_{\text{pred}} \gets h_\psi(\tilde{z})$
    \STATE $p_{\text{cluster}} \gets \text{softmax}(\text{ClusterHead}(z_{\text{online}}))$
    \STATE $\mathcal{L}_{\text{JEPA}} \gets \text{MSE}(z_{\text{pred}}[\mathcal{M}], z_{\text{target}}[\mathcal{M}])$
    \STATE $\mathcal{L}_{\text{cluster}} \gets \text{KL}(q_{\text{gmm}}[\mathcal{M}] \| p_{\text{cluster}}[\mathcal{M}])$
    \STATE $\lambda \gets \lambda_{\text{start}} + (\lambda_{\text{end}} - \lambda_{\text{start}}) \cdot t / T_{\text{max}}$
    \STATE $\mathcal{L} \gets \mathcal{L}_{\text{JEPA}} + \lambda \cdot \mathcal{L}_{\text{cluster}}$
    \STATE Update $\phi, \psi$ via $\nabla \mathcal{L}$
    \STATE $\phi' \gets \tau \phi' + (1-\tau)\phi$ \COMMENT{EMA update}
\ENDFOR
\end{algorithmic}
\end{algorithm}

\subsection{Architecture}

The encoder uses strided convolutions followed by a Conformer stack. Total stride $[8 \times 8 \times 5] = 320$ yields 20ms frames at 16kHz. Each strided block applies convolution, Snake-Beta activation \cite{ziyin2020snake}, residual dilated convolutions, and density-adaptive attention \cite{ioannides2024daam}. The Conformer stack uses 4 layers with gated relative position bias and layer aggregation via cross-attention.

\paragraph{Masking Strategy.} Block-wise temporal masking samples contiguous spans of length 10--25 frames until 40--65\% of positions are masked. A learned mask token $\mathbf{t}_{\text{mask}} \in \mathbb{R}^{C}$ replaces masked positions.

Architectural choices are shared across all models including baselines; GMM anchoring + JEPA loss versus original WavLM loss objective is the only difference between GMM-JEPA and WavLM-style. See Appendix~\ref{app:architecture} for full specifications.

\section{Experiments}

\paragraph{Setup.} Pre-trained on approximately 50,000 hours of speech from LibriLight large subset and English Granary \cite{bai2024granary}. To assess whether GMM anchoring generalizes across architectures, two encoder variants are evaluated: a Conformer-based encoder (GMM-JEPA) and a transformer-based encoder (GMM-JEPA-T). The transformer variant uses 10 standard transformer layers \cite{vaswani2017attention} with a 7-layer CNN frontend, enabling comparison across architectures. Both share identical GMM supervision and training procedures (Table~\ref{tab:hyperparams}). Baselines include Pure JEPA ($\lambda=0$) and a WavLM-style baseline that uses our architecture with k-means clustering on log-mel features. This provides a controlled comparison that isolates the effect of GMM soft anchoring; direct comparison to published WavLM results would be confounded by differences in model size, training data, and compute.

\paragraph{Evaluation.}
All downstream tasks train on frozen encoder representations. For ASR, a 2-layer BiLSTM with CTC loss \cite{graves2006ctc} is trained on LibriSpeech \texttt{train-clean-100} \cite{panayotov2015librispeech} with SpecAugment \cite{park2019specaugment}, using greedy decoding without a language model. For emotion recognition, a linear classifier is trained on mean-pooled representations using IEMOCAP \cite{busso2008iemocap} with 5-fold cross-validation over four emotions (angry, happy, sad, neutral). For slot filling, a sequence model is trained on SNIPS \cite{coucke2018snips} with speaker-disjoint splits.

\paragraph{Results.} Table~\ref{tab:main} shows ASR performance. GMM-JEPA-T achieves 28.68\% WER compared to 33.22\% for WavLM-style. Pure JEPA produces 100\% WER, confirming complete representation collapse \cite{jing2022understanding, sansone2024collapse, mo2024connecting} without acoustic grounding.

Table~\ref{tab:sf} shows slot filling results. GMM-JEPA achieves the highest Slot Type F1 (64.7\%), outperforming WavLM-style by 5.6\% absolute.

Table~\ref{tab:ser_folds} shows emotion recognition results. GMM-JEPA-T and GMM-JEPA achieve 67.76\% and 67.30\% average accuracy respectively versus 65.46\% for WavLM-style, with consistent improvement across all folds.

\begin{table}[t]
  \caption{\textbf{GMM-JEPA reduces WER by 14\% relative.} ASR on LibriSpeech \texttt{dev-clean}. Pure JEPA collapses (100\% WER).}
  \label{tab:main}
  \begin{center}
    \begin{small}
      \begin{sc}
        \begin{tabular}{lccc}
          \toprule
          Model & WER (\%) $\downarrow$ & CER (\%) $\downarrow$ & $\Delta$ WER \\
          \midrule
          Pure JEPA & 100.00 & 93.11 & -- \\
          WavLM-style & 33.22 & 11.28 & \textit{baseline} \\
          \midrule
          \rowcolor{ourscolor} GMM-JEPA & 29.18 & 9.62 & $-$12.2\% \\
          \rowcolor{ourscolor} GMM-JEPA-T & \textbf{28.68} & \textbf{9.44} & $-$13.7\% \\
          \bottomrule
        \end{tabular}
      \end{sc}
    \end{small}
  \end{center}
\end{table}

\begin{table}[t]
  \caption{\textbf{GMM-JEPA outperforms WavLM-style by 5.6\% absolute.} Slot filling on SNIPS.}
  \label{tab:sf}
  \begin{center}
    \begin{small}
      \begin{sc}
        \begin{tabular}{lccc}
          \toprule
          Model & Type F1 $\uparrow$ & Edit F1 $\uparrow$ & $\Delta$ Type \\
          \midrule
          Pure JEPA & 5.0 & 2.2 & -- \\
          WavLM-style & 59.1 & 33.7 & \textit{baseline} \\
          \midrule
          \rowcolor{ourscolor} GMM-JEPA-T & 59.2 & 32.8 & $+$0.1 \\
          \rowcolor{ourscolor} GMM-JEPA & \textbf{64.7} & \textbf{36.0} & $+$5.6 \\
          \bottomrule
        \end{tabular}
      \end{sc}
    \end{small}
  \end{center}
\end{table}

\begin{table}[t]
  \caption{\textbf{GMM-JEPA improves emotion recognition by 2\% absolute.} SER accuracy (\%) consistent across IEMOCAP folds.}
  \label{tab:ser_folds}
  \begin{center}
    \begin{tiny}
      \begin{sc}
        \begin{tabular}{lcccccc|c}
          \toprule
          Model & F1 & F2 & F3 & F4 & F5 & Avg & $\Delta$ \\
          \midrule
          Pure JEPA & 48.88 & 46.12 & 49.89 & 48.67 & 46.97 & 48.11 & -- \\
          WavLM-style & 64.83 & 64.19 & 65.30 & 67.11 & 65.85 & 65.46 & \textit{base} \\
          \midrule
          \rowcolor{ourscolor} GMM-JEPA & 66.07 & \textbf{67.85} & 66.67 & \textbf{68.56} & 67.37 & 67.30 & $+$1.8 \\
          \rowcolor{ourscolor} GMM-JEPA-T & \textbf{67.19} & 67.52 & \textbf{67.58} & 68.11 & \textbf{68.41} & \textbf{67.76} & $+$2.3 \\
          \bottomrule
        \end{tabular}
      \end{sc}
    \end{tiny}
  \end{center}
\end{table}

\section{Analysis}
\label{sec:analysis}

To understand \emph{why} GMM anchoring helps, the learned representations are analyzed through clustering and visualization.

\paragraph{Evaluation Methodology.}
Frame-level representations are extracted from each encoder on 1,000 utterances (up to 10 seconds each). Cluster assignments are obtained directly from each model's learned cluster prediction head ($K=1024$ classes), and two metrics are computed:

\textbf{Cluster entropy} measures vocabulary utilization:
\begin{equation}
H = -\sum_{k=1}^{K} p_k \log p_k, \quad p_k = \frac{|\{t : c_t = k\}|}{T_{\text{total}}},
\end{equation}
normalized by $\log K$ to yield a percentage. High entropy indicates all clusters are actively used; low entropy suggests collapse to few dominant clusters.

\textbf{Adjacent consistency} measures temporal smoothness:
\begin{equation}
\text{Consistency} = \frac{1}{T-1} \sum_{t=1}^{T-1} \mathbf{1}[c_t = c_{t+1}].
\end{equation}
Since phonemes typically span 50--200ms \cite{ladefoged2014course} while frames are 20ms, acoustically meaningful clusters should persist across adjacent frames rather than flickering rapidly.

For visualization, 10,000 randomly sampled frames are projected to 2D via UMAP \cite{mcinnes2018umap} and colored by their predicted cluster assignment (top 10 clusters highlighted).

\paragraph{Results.}
Table~\ref{tab:metrics} summarizes the quantitative findings.

\begin{table}[t]
  \caption{\textbf{GMM-JEPA achieves 98\% entropy vs. 31\% for WavLM-style.} Cluster analysis metrics with $K=1024$.}
  \label{tab:metrics}
  \begin{center}
    \begin{small}
      \begin{sc}
        \begin{tabular}{lccc}
          \toprule
          Model & Entropy $\uparrow$ & Used $\uparrow$ & Adj. Cons. \\
          \midrule
          Pure JEPA & 45\% & 516 & 0.205 \\
          WavLM-style & 31\% & 978 & 0.750 \\
          \midrule
          \rowcolor{ourscolor} GMM-JEPA & 85\% & 1007 & 0.395 \\
          \rowcolor{ourscolor} GMM-JEPA-T & \textbf{98\%} & \textbf{1013} & 0.299 \\
          \bottomrule
        \end{tabular}
      \end{sc}
    \end{small}
  \end{center}
\end{table}

The results reveal clear differences between methods. Pure JEPA achieves only 45\% entropy using 516/1024 clusters, indicating severe representation collapse \cite{jing2022understanding} where half the cluster vocabulary is unused. WavLM-style performs even worse in terms of entropy (31\%) despite using more clusters (978/1024), indicating extreme imbalance where a few clusters dominate while most are rarely assigned, a phenomenon observed in other self-supervised methods \cite{caron2020unsupervised}. 

In contrast, both GMM-anchored variants achieve substantially higher entropy. GMM-JEPA reaches 85\% entropy with 1007/1024 clusters used, while GMM-JEPA-T achieves the best result at 98\% entropy with 1013/1024 clusters. This demonstrates that GMM anchoring effectively prevents the representation collapse observed in both Pure JEPA and WavLM-style training.

\paragraph{Qualitative Analysis.}
Figure~\ref{fig:umap} visualizes the representation space via UMAP projection. Pure JEPA collapses to a small, dense region with limited cluster diversity. WavLM-style spreads across a larger area but exhibits diffuse, overlapping cluster assignments with no clear separation. GMM-JEPA-T and GMM-JEPA both form localized, well-separated cluster regions distributed across the embedding space, demonstrating cleaner partitioning of the latent space.

Figure~\ref{fig:dist} confirms the entropy findings through cluster utilization distributions. Pure JEPA shows a steep dropoff after approximately 270 clusters, with the remaining clusters receiving near-zero assignments. WavLM-style also exhibits a steeper dropoff despite broader cluster usage, explaining its low entropy. Both GMM-anchored variants maintain substantially flatter distributions, with GMM-JEPA-T showing the most uniform utilization across all 1024 clusters. Figure~\ref{fig:utterance} examines cluster assignment behavior over time. Pure JEPA shows near-zero prediction confidence with rapid cluster flickering, indicating degenerate representations. WavLM-style collapses to a smaller set of dominant clusters with moderate confidence, explaining its low entropy despite using many clusters. In contrast, both GMM-JEPA variants maintain higher confidence than their WavLM and Pure JEPA counterparts with temporally coherent cluster spans, demonstrating that GMM anchoring produces stable, meaningful cluster assignments.

\subsection{Embedding Quality Analysis}

\paragraph{Cluster Quality Analysis.}
To assess representational similarity across models, embeddings are clustered via K-means and compared using Normalized Mutual Information (NMI). NMI measures the agreement between two clusterings, defined as:
\begin{equation}
\text{NMI}(U, V) = \frac{2 \cdot I(U; V)}{H(U) + H(V)},
\end{equation}
where $I(U; V)$ is the mutual information between clusterings $U$ and $V$, and $H(\cdot)$ denotes entropy. NMI ranges from 0 (independent) to 1 (identical).

\begin{figure*}[t]
\centering
\begin{subfigure}[t]{0.32\textwidth}
    \includegraphics[width=\textwidth]{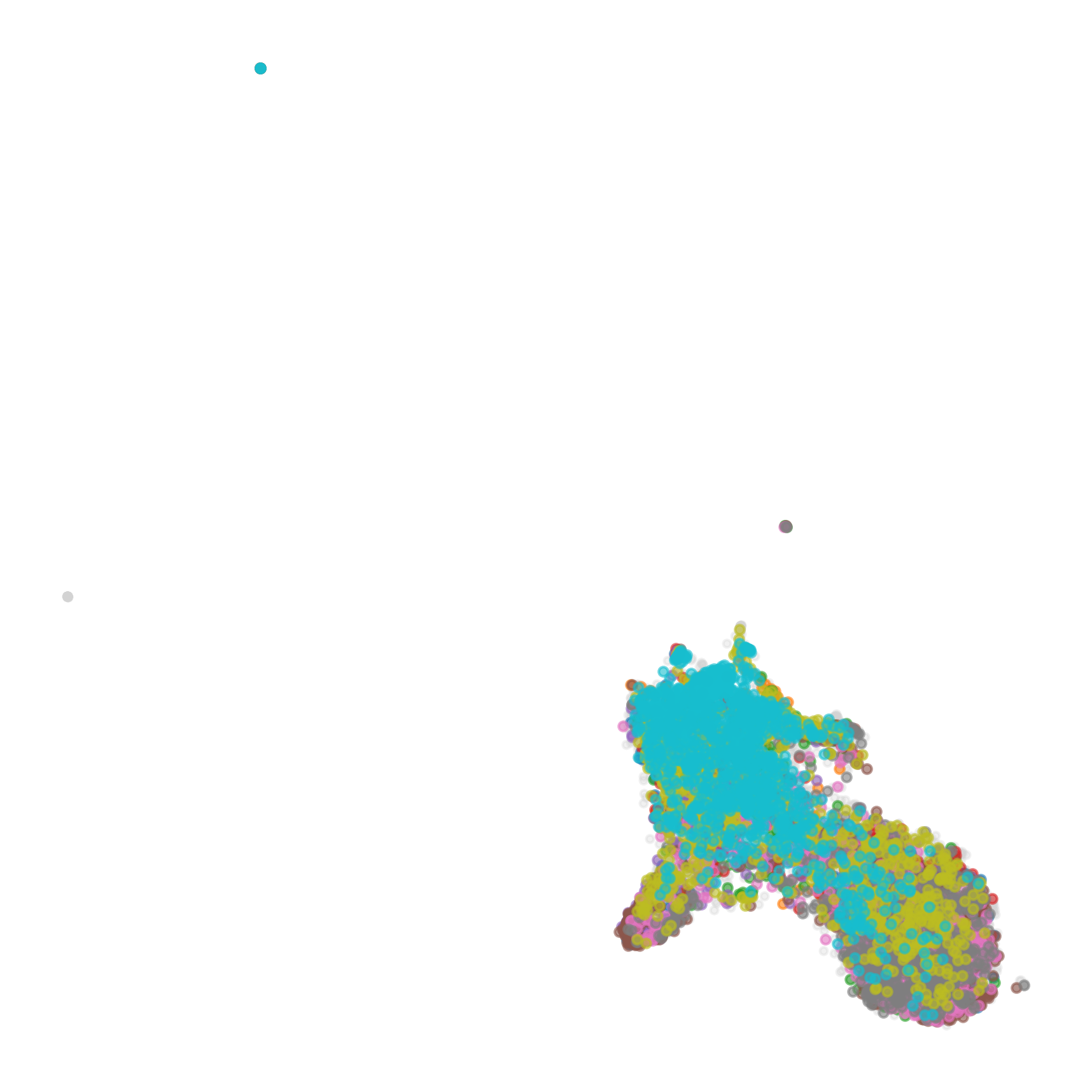}
    \caption{Diffuse clusters}
\end{subfigure}
\hfill
\begin{subfigure}[t]{0.32\textwidth}
    \includegraphics[width=\textwidth]{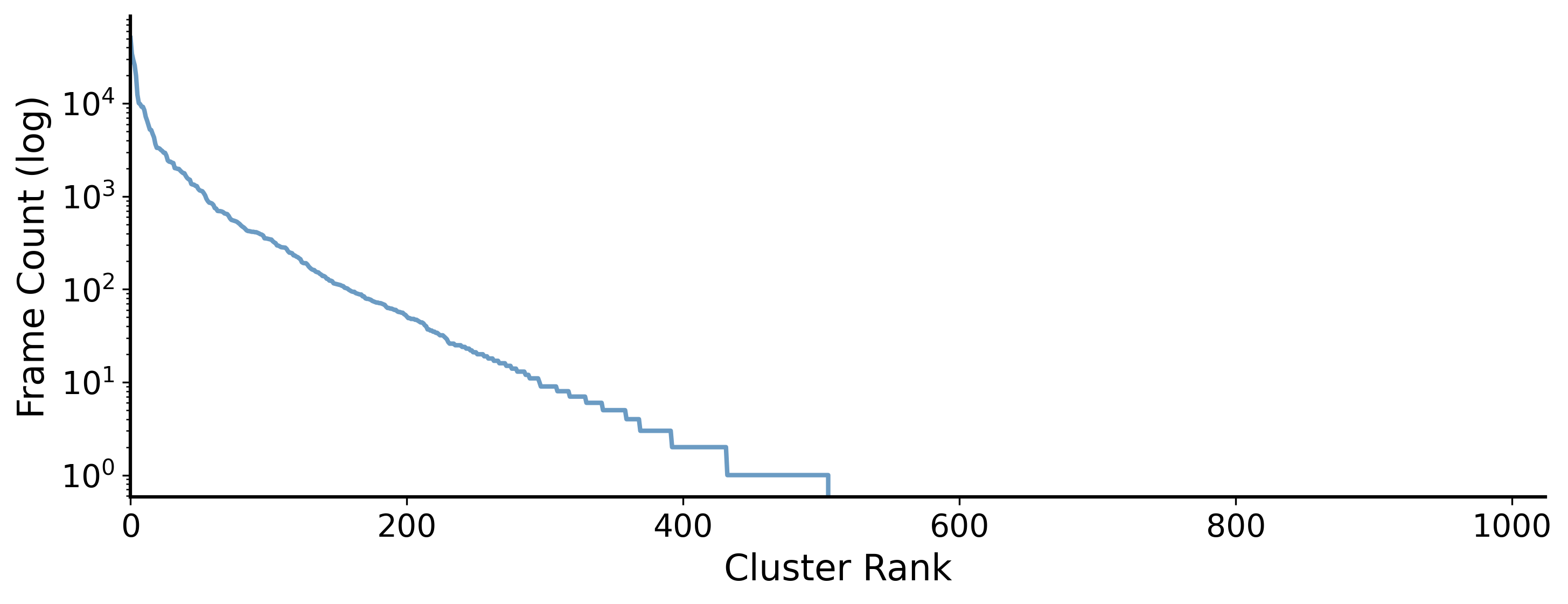}
    \caption{55.9\% entropy}
\end{subfigure}
\hfill
\begin{subfigure}[t]{0.32\textwidth}
    \includegraphics[width=\textwidth]{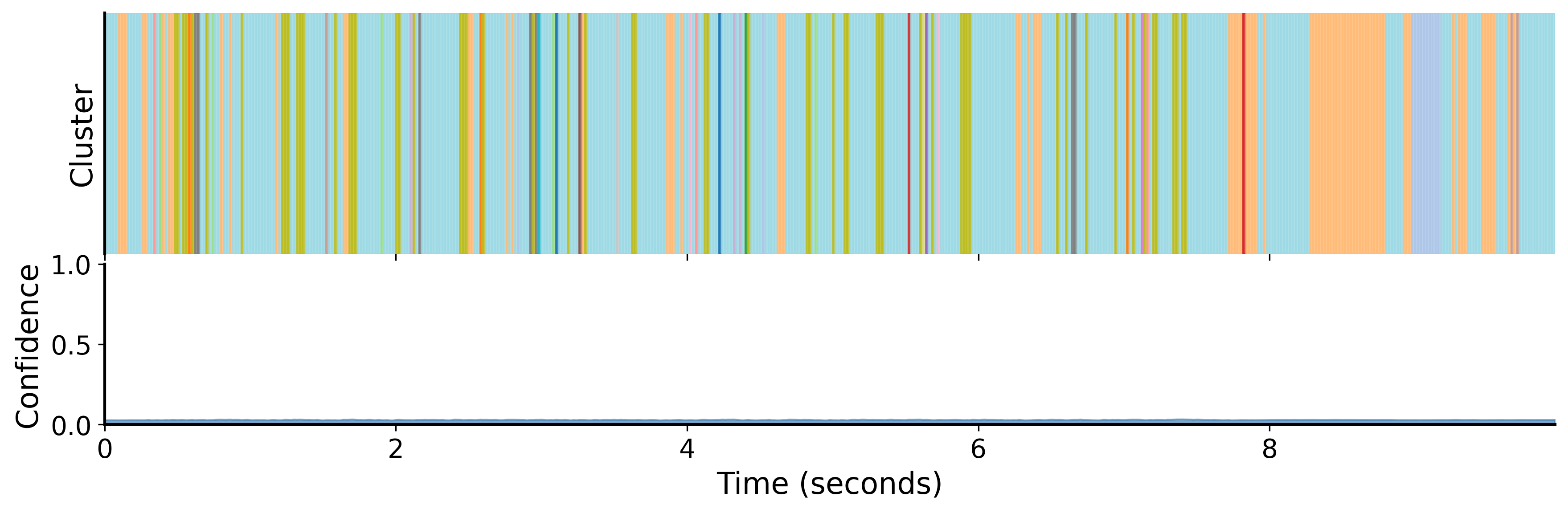}
    \caption{Unstable assignments}
\end{subfigure}
\caption{\textbf{Removing residual GMM supervision ($\lambda_{\text{end}}=0$) causes collapse.} Without ongoing anchoring, representations deteriorate. \textbf{(a)} UMAP shows overlapping clusters. \textbf{(b)} Only 506/1024 clusters used. \textbf{(c)} Confidence drops, flickering increases.}
\label{fig:ablation}
\end{figure*}

\paragraph{Linear Probe Evaluation.}
Linear probes are trained to assess how well frozen embeddings classify both high-level paralinguistic categories (Speaker ID, Gender) and low-level linguistic units (phonemes). All models are trained for the same number of speech hours to ensure fair comparison. Results are shown in Table~\ref{tab:objective_comparison} and Table~\ref{tab:phoneme_probe}.

\begin{table}[t]
  \caption{\textbf{All anchored models saturate on speaker/gender; Pure JEPA fails.} Linear probe accuracy (\%) on paralinguistic categories.}
  \label{tab:objective_comparison}
  \begin{center}
    \begin{small}
      \begin{sc}
        \begin{tabular}{lcc}
          \toprule
          Model & Speaker $\uparrow$ & Gender $\uparrow$ \\
          \midrule
          Pure JEPA & 18.7{\scriptsize $\pm$1.8} & 69.2{\scriptsize $\pm$1.5} \\
          WavLM-style & 99.6{\scriptsize $\pm$0.3} & 99.4{\scriptsize $\pm$0.2} \\
          \midrule
          \rowcolor{ourscolor} GMM-JEPA-T & 99.7{\scriptsize $\pm$0.2} & 99.7{\scriptsize $\pm$0.2} \\
          \rowcolor{ourscolor} GMM-JEPA & \textbf{99.8}{\scriptsize $\pm$0.2} & \textbf{99.6}{\scriptsize $\pm$0.3} \\
          \bottomrule
        \end{tabular}
      \end{sc}
    \end{small}
  \end{center}
\end{table}

\begin{table}[t]
  \caption{\textbf{GMM-JEPA improves phoneme classification by +5.4\% (linear) and +6.7\% (MLP).} Phoneme probe accuracy (\%).}
  \label{tab:phoneme_probe}
  \begin{center}
    \begin{small}
      \begin{sc}
        \begin{tabular}{lcc}
          \toprule
          Model & Linear $\uparrow$ & MLP $\uparrow$ \\
          \midrule
          Pure JEPA & 7.4{\scriptsize $\pm$0.0} & 9.0{\scriptsize $\pm$0.1} \\
          WavLM-style & 47.6{\scriptsize $\pm$0.2} & 57.5{\scriptsize $\pm$0.2} \\
          \midrule
          \rowcolor{ourscolor} GMM-JEPA-T & 51.4{\scriptsize $\pm$0.1} & 59.4{\scriptsize $\pm$0.4} \\
          \rowcolor{ourscolor} GMM-JEPA & \textbf{53.0}{\scriptsize $\pm$0.1} & \textbf{64.2}{\scriptsize $\pm$0.2} \\
          \bottomrule
        \end{tabular}
      \end{sc}
    \end{small}
  \end{center}
\end{table}

Pure JEPA fails catastrophically across all metrics (18.7\% speaker, 7.4\% phoneme), confirming that acoustic grounding is essential for learning meaningful representations. GMM-JEPA consistently matches or outperforms WavLM-style, with notable improvements in phoneme classification (+5.4\% linear, +6.7\% MLP). GMM-JEPA-T achieves comparable performance, demonstrating that the benefits of GMM anchoring transfer across architectures. Table~\ref{tab:clustering_comparison} shows NMI between embedding clusters.

\begin{table}[t]
  \caption{\textbf{Pure JEPA shares no structure with other models (NMI$<$0.11).} NMI between embedding clusters.}
  \label{tab:clustering_comparison}
  \begin{center}
    \begin{small}
      \begin{sc}
        \begin{tabular}{lcccc}
          \toprule
          & Pure & WavLM & GMM & GMM-T \\
          \midrule
          Pure JEPA & 1.00 & 0.08 & 0.11 & 0.06 \\
          WavLM-style & 0.08 & 1.00 & 0.47 & 0.57 \\
          \midrule
          \rowcolor{ourscolor} GMM-JEPA & 0.11 & 0.47 & 1.00 & 0.47 \\
          \rowcolor{ourscolor} GMM-JEPA-T & 0.06 & 0.57 & 0.47 & 1.00 \\
          \bottomrule
        \end{tabular}
      \end{sc}
    \end{small}
  \end{center}
\end{table}

Pure JEPA shows minimal similarity to all other models (NMI$<$0.11), consistent with the representation collapse observed in Section~\ref{sec:analysis}. The three successful models exhibit moderate mutual similarity (NMI=0.47--0.57), indicating they capture related but distinct acoustic structure. GMM-JEPA and GMM-JEPA-T show medium similarity (NMI=0.47), suggesting that architecture influences representation geometry independently of training objective.

\paragraph{Cluster-Label Alignment.}
Finally, the alignment between unsupervised clusters and supervised labels is measured via NMI (Table~\ref{tab:supervised_nmi}).

\begin{table}[t]
  \caption{\textbf{GMM-JEPA captures speaker/gender; GMM-JEPA-T captures phonemes.} Cluster-label alignment (NMI).}
  \label{tab:supervised_nmi}
  \begin{center}
    \begin{small}
      \begin{sc}
        \begin{tabular}{lccc}
          \toprule
          Model & Phoneme $\uparrow$ & Speaker $\uparrow$ & Gender $\uparrow$ \\
          \midrule
          Pure JEPA & 0.03 & 0.14 & 0.02 \\
          WavLM-style & 0.20 & 0.27 & 0.10 \\
          \midrule
          \rowcolor{ourscolor} GMM-JEPA & 0.19 & \textbf{0.48} & \textbf{0.15} \\
          \rowcolor{ourscolor} GMM-JEPA-T & \textbf{0.24} & 0.22 & 0.10 \\
          \bottomrule
        \end{tabular}
      \end{sc}
    \end{small}
  \end{center}
\end{table}

GMM-JEPA achieves the highest alignment for Speaker (0.48) and Gender (0.15), substantially outperforming WavLM-style on paralinguistic categories. GMM-JEPA-T achieves the highest phoneme alignment (0.24), suggesting the transformer architecture better preserves fine-grained linguistic structure. The Conformer architecture appears to emphasize paralinguistic structure while the Transformer preserves finer phonetic detail; this architectural difference is orthogonal to GMM anchoring, which improves both variants over their respective baselines. These results indicate that GMM anchoring encourages representations that capture speaker-level information more effectively, which may explain the improved emotion recognition performance observed in Table~\ref{tab:ser_folds}.

\subsection{Ablation: GMM Anchoring}

A natural question is whether GMM supervision can be fully removed after the initial grounding phase. To investigate, a GMM-JEPA model is trained with $\lambda$ decaying to zero rather than to 0.01 (i.e., disabling GMM loss in late training).
\begin{table}[t]
  \caption{\textbf{Anchoring is essential: $\lambda_{\text{end}}=0$ degrades WER to 40.9\%.} Effect of removing GMM supervision after decay.}
  \label{tab:lambda_ablation}
  \begin{center}
    \resizebox{\columnwidth}{!}{%
        \begin{tabular}{lcccccc}
          \toprule
          \textsc{$\lambda_{\text{end}}$} & \textsc{Ent.}$\uparrow$ & \textsc{Used}$\uparrow$ & \textsc{WER} $\downarrow$ & \textsc{SER} $\uparrow$ & \textsc{SF-T} $\uparrow$ & \textsc{SF-E} $\uparrow$ \\
          \midrule
          0.00 & 57.7 & 530 & 40.9 & 63.8 & 62.3 & 34.2 \\
          \rowcolor{ourscolor} 0.01 & \textbf{84.7} & \textbf{1011} & \textbf{29.2} & \textbf{67.3} & \textbf{64.7} & \textbf{36.0} \\
          \bottomrule
        \end{tabular}
    }
  \end{center}
\end{table}

As shown in Table~\ref{tab:lambda_ablation}, removing residual GMM supervision causes severe deterioration across both representation quality and downstream performance. Entropy drops from 84.7\% to 57.7\%, and cluster utilization falls from 1011 to 506 clusters, approximately the same as Pure JEPA's collapsed state (45\% entropy, 516 clusters). Critically, this degradation translates directly to downstream tasks: WER increases from 29.18\% to 40.95\%, SER accuracy drops from 67.30\% to 63.79\%, and slot filling performance decreases across both Type F1 (64.7\% to 62.3\%) and Edit F1 (36.0\% to 34.2\%). 

Figure~\ref{fig:ablation} confirms this degradation visually. The UMAP projection (Figure~\ref{fig:ablation}a) shows that without residual anchoring, clusters become diffuse and overlapping, losing the well-separated structure observed in GMM-JEPA. The cluster distribution (Figure~\ref{fig:ablation}b) exhibits a steep dropoff, with nearly half the clusters receiving negligible assignments. The confidence plot (Figure~\ref{fig:ablation}c) shows reduced confidence with increased cluster flickering compared to the stable, high-confidence assignments of GMM-JEPA ($\lambda_{\text{end}}=0.01$).

These results demonstrate that GMM anchoring provides ongoing regularization, not merely initialization. Even minimal supervision ($\lambda=0.01$) prevents the representational drift that occurs when the JEPA objective dominates unconstrained, and this regularization is essential for maintaining downstream task performance.

\section{Discussion}

Our results demonstrate that frozen GMM supervision addresses a core challenge in JEPA-based speech SSL: without external grounding, the EMA teacher provides no spectral information, allowing representations to collapse to acoustically meaningless states. The GMM provides this grounding through two mechanisms. First, early training ($\lambda=1.0$) forces the encoder to predict acoustic cluster posteriors, establishing phonetically meaningful structure before the JEPA objective can cause collapse. Second, residual supervision ($\lambda=0.01$) maintains this structure throughout training, as our ablation shows that removing the anchor causes representations to deteriorate even after initial grounding.

We hypothesize that the soft assignment property of GMMs provides an advantage over hard k-means clustering. At acoustic boundaries where frames are ambiguous between phonetic categories, GMM posteriors preserve this uncertainty rather than forcing arbitrary hard decisions. This may explain why GMM-JEPA achieves higher cluster entropy (85--98\%) compared to WavLM-style (31\%), which uses hard assignments that concentrate probability mass on dominant clusters. Validating this hypothesis by comparing soft GMM posteriors against hard GMM assignments remains future work.

\paragraph{Limitations \& Future Work.}
Our models are trained on 50k hours in a single pass, while HuBERT and WavLM use iterative re-clustering with multiple passes, effectively multiplying training compute. We focus on demonstrating that frozen GMM supervision provides effective acoustic grounding for JEPA rather than achieving state-of-the-art absolute performance; we do not compare against published HuBERT/WavLM checkpoints due to confounds in architecture and training data. We do not ablate soft versus hard GMM assignments, leaving validation of the soft assignment hypothesis for future work. The optimal decay schedule and residual weight ($\lambda_{\text{end}}=0.01$) were selected based on preliminary experiments but not extensively tuned. Additionally, we evaluate on English speech only.

Promising directions include: (1) allowing the GMM to adapt during training rather than remaining frozen, potentially capturing evolving acoustic structure; (2) exploring alternative anchors such as neural density estimators; (3) extending to multilingual settings where iterative re-clustering becomes computationally prohibitive; and (4) investigating whether GMM anchoring benefits other JEPA domains beyond speech.

\section{Conclusion}

We introduced GMM-Anchored JEPA, a framework that grounds self-supervised speech representations with frozen Gaussian Mixture Model supervision. By fitting a GMM once on log-mel spectrograms and using its soft posteriors as auxiliary targets with a decaying weight schedule, we eliminate the need for iterative re-clustering while preserving uncertainty at acoustic boundaries. Experiments on 50k hours of speech show that GMM anchoring improves ASR, emotion recognition, and slot filling compared to WavLM-style baselines with matched compute, while achieving substantially higher cluster entropy (98\% vs.\ 31\%). Ablation studies confirm that residual GMM supervision provides ongoing regularization essential for preventing representation collapse. Our results suggest that simple, frozen acoustic anchors can effectively stabilize JEPA training for speech, opening new directions for efficient self-supervised speech representation learning.

\section*{Impact Statement}
This work advances self-supervised speech representation learning, with applications in ASR, emotion recognition, and spoken language understanding. Improved representations could benefit accessibility tools and low-resource languages. However, these representations could potentially be misused for surveillance or voice profiling. We encourage responsible deployment with appropriate consent and privacy protections.

\bibliographystyle{plain}
\bibliography{references}

\newpage
\text{}
\newpage
\onecolumn
\appendix

\section{Architecture Details}
\label{app:architecture}

\subsection{Encoder Backbone}

The encoder consists of strided convolutional blocks followed by a Conformer stack.

\paragraph{Input Convolution.} A 7-kernel convolution maps raw waveform to initial features:
\begin{equation}
h_0 = \text{Conv1d}^{k=7,\,p=3}_{1 \to C_0}(\mathbf{x}).
\end{equation}

\paragraph{Strided Encoder Blocks.} Each block $i$ with stride $s_i$ applies:
\begin{enumerate}
    \item Strided convolution with kernel $2s_i$, stride $s_i$, padding $s_i/2$
    \item Snake-Beta activation (see below)
    \item Residual blocks with dilated convolutions (dilation pattern $[1, 3, 5]$ per block)
    \item Density Adaptive Attention block
\end{enumerate}

Channel progression $[32, 64, 128, 256]$ with strides $[8, 8, 5]$ yields total stride 320.

\paragraph{Conformer Stack.} $N_{\text{conf}}=4$ Conformer layers with:
\begin{itemize}
    \item Half-step feed-forward modules (expansion factor 4)
    \item Multi-head self-attention (32 heads) with gated relative position bias
    \item Depthwise separable convolution module with GLU activation \cite{dauphin2017glu} (kernel size 31)
\end{itemize}

Each Conformer block applies:
\begin{align}
x &= x + \frac{1}{2}\text{FFN}_1(x), \\
x &= x + \text{MHSA}(x), \\
x &= x + \text{ConvModule}(x), \\
x &= x + \frac{1}{2}\text{FFN}_2(x).
\end{align}

\subsection{Snake-Beta Activation}

\begin{equation}
\text{Snake}_\alpha(x) = x + \frac{\sin^2(\alpha x)}{\alpha}, \quad \alpha = \text{softplus}(a) + 0.01,
\end{equation}
where $a \in \mathbb{R}^{1 \times C \times 1}$ is a learnable parameter per channel.

\subsection{Gated Relative Position Bias}

\paragraph{Position Embedding.} Relative positions are mapped to buckets via logarithmic bucketing:
\begin{equation}
\text{bucket}(i-j) = \begin{cases}
|i-j| & \text{if } |i-j| < \frac{B}{4} \\
\frac{B}{4} + \frac{B}{4} \cdot \frac{\log(|i-j|/\frac{B}{4})}{\log(D_{\max}/\frac{B}{4})} & \text{otherwise}
\end{cases}
\end{equation}
where $B=320$ buckets and $D_{\max}=800$ maximum distance.

\paragraph{Gating Mechanism.} Per-position gates computed from queries $\mathbf{q} \in \mathbb{R}^{B \times H \times T \times d}$:
\begin{align}
g_{\text{update}} &= \sigma(\mathbf{q} \cdot \mathbf{u}), \\
g_{\text{reset}} &= \sigma(\mathbf{q} \cdot \mathbf{w}),
\end{align}
where $\mathbf{u}, \mathbf{w} \in \mathbb{R}^{H \times d}$ are learnable vectors.

\paragraph{Gated Bias.} Final position bias combines learned embeddings with gating:
\begin{align}
\tilde{r}_{i-j} &= s \cdot g_{\text{reset},i} \cdot d_{i-j}, \\
r_{i-j} &= d_{i-j} + g_{\text{update},i} \cdot d_{i-j} + (1 - g_{\text{update},i}) \cdot \tilde{r}_{i-j},
\end{align}
where $d_{i-j}$ is the learned embedding and $s$ is a learnable scale.

\subsection{Layer Aggregation}

Given Conformer layer outputs $\{z^{(l)}\}_{l=0}^{L}$:
\begin{align}
\mathbf{Q} &= W_Q \cdot \text{pool}(z^{(L)}), \\
\mathbf{K} &= W_K \cdot \text{stack}(\text{pool}(z^{(l)})), \\
\mathbf{V} &= W_V \cdot \text{stack}(\text{pool}(z^{(l)})), \\
\text{weights} &= \text{softmax}(\mathbf{Q} \mathbf{K}^\top / \sqrt{d}), \\
z_{\text{agg}} &= \sum_l \text{weights}_l \cdot z^{(l)},
\end{align}
where pooling is over the time dimension and attention operates over layers.

\subsection{Density Adaptive Attention (DAAM)}

DAAM \cite{ioannides2024daam, ioannides2024daam_speech} computes attention weights using a product of Gaussian kernels over normalized activations. For input $x \in \mathbb{R}^{B \times C \times T}$:

\paragraph{Normalization.} Input is standardized along the attention axis:
\begin{equation}
\bar{x} = \frac{x - \mu_x}{\sigma_x + \epsilon}.
\end{equation}

\paragraph{Gaussian Mixture Weights.} For $N_g$ Gaussians with learnable offsets $\delta_i$ and scales $c_i$:
\begin{align}
\log w_i &= -\frac{(\bar{x} - \delta_i)^2}{2c_i^2} - \frac{1}{2}\log(2\pi c_i^2), \\
\log w &= \sum_{i=1}^{N_g} \log w_i.
\end{align}

\paragraph{Output.} Attention is applied with residual connection:
\begin{equation}
\text{out} = x \odot \text{softmax}(\log w) + x.
\end{equation}

The input is split along the channel dimension into $H$ heads, each processed by an independent DAAM module, then concatenated.

\subsection{Cluster Prediction Head}

A multi-layer MLP with residual connections maps encoder outputs to $K$ cluster logits:
\begin{align}
h_0 &= \text{GELU}(\text{LayerNorm}(\text{Linear}_{C \to H}(z))), \\
h_l &= h_{l-1} + \text{Block}_l(h_{l-1}), \quad l = 1, \ldots, L_{\text{head}}, \\
\text{logits} &= \text{Linear}_{H \to K}(\text{LayerNorm}(h_{L_{\text{head}}})),
\end{align}
where each block contains LayerNorm $\to$ Linear $\to$ GELU $\to$ Dropout $\to$ Linear $\to$ Dropout with residual.
\subsection{Predictor Architecture}
The predictor consists of: (1) Conv1d projection: $C \to C$, (2) GELU activation, (3) Single Conformer block (shared relative position bias with encoder), and (4) Conv1d output projection: $C \to C$.

\subsection{Target Encoder and EMA Update}

The target encoder has identical architecture but parameters are updated via EMA after each gradient step:
\begin{equation}
\phi' \leftarrow \tau \phi' + (1 - \tau) \phi, \quad \tau = 0.996.
\end{equation}

\section{Training Hyperparameters}
\label{app:hyperparams}

\begin{table}[H]
  \caption{Training hyperparameters.}
  \label{tab:hyperparams}
  \begin{center}
    \begin{small}
      \begin{sc}
        \begin{tabular}{lll}
          \toprule
          & Conformer & Transformer \\
          \midrule
          \multicolumn{3}{l}{\textit{Encoder Architecture}} \\
          Parameters & 31.1M & 41.3M \\
          CNN layers & 3 & 7 \\
          Channels & [32,64,128,256] & [256]$\times$7 \\
          Strides & [8,8,5] & [5,2,2,2,2,2,2] \\
          Latent dim & 512 & 512 \\
          Encoder layers & 4 & 10 \\
          Attention heads & 32 & 8 \\
          FFN mult / dim & 4$\times$512 & 2048 \\
          Conv kernel & 31 & -- \\
          \midrule
          \multicolumn{3}{l}{\textit{GMM Fitting}} \\
          Components $K$ & \multicolumn{2}{c}{1024} \\
          Mel bins & \multicolumn{2}{c}{80} \\
          Initialization & \multicolumn{2}{c}{K-means++} \\
          Optimization & \multicolumn{2}{c}{Mini-batch SGD} \\
          Learning rate & \multicolumn{2}{c}{$10^{-2} \to 10^{-4}$ (cosine)} \\
          Training frames & \multicolumn{2}{c}{9.6B} \\
          \midrule
          \multicolumn{3}{l}{\textit{K-Means Fitting (WavLM-style baseline)}} \\
          Components $K$ & \multicolumn{2}{c}{1024} \\
          Feature type & \multicolumn{2}{c}{Log-mel (80 bins)} \\
          Initialization & \multicolumn{2}{c}{K-means++} \\
          Update method & \multicolumn{2}{c}{Lloyd's algorithm} \\
          Lloyd iterations & \multicolumn{2}{c}{20} \\
          Training frames & \multicolumn{2}{c}{9.6B} \\
          \midrule
          \multicolumn{3}{l}{\textit{Training}} \\
          Optimizer & \multicolumn{2}{c}{AdamW \cite{loshchilov2019adamw}} \\
          Learning rate & \multicolumn{2}{c}{$10^{-5} \to 10^{-4} \to 10^{-5}$} \\
          LR schedule & \multicolumn{2}{c}{$10\%$ linear warmup / decay} \\
          Weight decay & \multicolumn{2}{c}{$10^{-3}$} \\
          Micro Batch size & 96 (Conformer) & 192 (Transf.) \\
          Gradient clipping & \multicolumn{2}{c}{1.0} \\
          GPUs & \multicolumn{2}{c}{8 $\times$ B200} \\
          EMA $\tau$ & \multicolumn{2}{c}{0.996} \\
          $\lambda$ decay & \multicolumn{2}{c}{$1.0 \to 0.01$} \\
          Mask ratio & \multicolumn{2}{c}{$[0.4, 0.65]$} \\
          Span length & \multicolumn{2}{c}{$[10, 25]$ frames} \\
          \midrule
          \multicolumn{3}{l}{\textit{Augmentation}} \\
          Noise SNR & \multicolumn{2}{c}{$[-5, 20]$ dB} \\
          Mix ratio & \multicolumn{2}{c}{$[-5, 5]$ dB} \\
          Noise/mix prob. & \multicolumn{2}{c}{0.25 each} \\
          \bottomrule
        \end{tabular}
      \end{sc}
    \end{small}
  \end{center}
\end{table}

\section{GMM Details}
\label{app:gmm}

\subsection{Posterior Computation}

The posterior with diagonal covariance is computed as:
\begin{equation}
q_k(\mathbf{m}) = \text{softmax}_k\left(\log \pi_k - \frac{D}{2}\log(2\pi) - \frac{1}{2}\sum_{d=1}^{D}\left[\log \sigma_{k,d}^2 + \frac{(m_d - \mu_{k,d})^2}{\sigma_{k,d}^2}\right]\right).
\end{equation}

\subsection{Chunked Soft Assignment}

To prevent out-of-memory errors with large $K$ and batch sizes, soft assignment is computed in chunks:

\begin{algorithm}[H]
\caption{Chunked Soft Assignment}
\begin{algorithmic}[1]
\STATE \textbf{Input:} Features $X \in \mathbb{R}^{N \times D}$, GMM parameters, chunk sizes
\STATE \textbf{Output:} Posteriors $Q \in \mathbb{R}^{N \times K}$
\FOR{$n = 0$ to $N$ step $\text{batch\_size}$}
    \FOR{$k = 0$ to $K$ step $\text{chunk\_k}$}
        \STATE $\Delta \gets X[n:n'] - \mu[k:k']$ \COMMENT{$[B, K', D]$}
        \STATE $\text{mahal} \gets \sum_d \Delta_d^2 / \sigma_{k,d}^2$
        \STATE $\log p_{n,k} \gets -\frac{D}{2}\log(2\pi) - \frac{1}{2}\sum_d \log \sigma_{k,d}^2 - \frac{1}{2}\text{mahal}$
    \ENDFOR
    \STATE $Q[n:n'] \gets \text{softmax}(\log \pi + \log p)$
\ENDFOR
\end{algorithmic}
\end{algorithm}

\section{Augmentation Details}
\label{app:augment}

\subsection{Energy-Based SNR Mixing}

Given clean signal $x$ and noise $n$, mixing at target SNR:
\begin{align}
E_x &= \frac{1}{|x|}\sum_i x_i^2, \quad E_n = \frac{1}{|n|}\sum_i n_i^2, \\
\alpha &= \sqrt{\frac{E_x}{10^{\text{SNR}/10} \cdot E_n}}, \\
x_{\text{aug}} &= x + \alpha \cdot n,
\end{align}
where SNR is sampled from $[-5, 20]$ dB with probability 0.25.

\subsection{Utterance Mixing}

A segment from another utterance is mixed with the primary:
\begin{enumerate}
    \item Sample mix length: $L_{\text{mix}} \gets \text{randint}(1, |x_1| \cdot 0.5)$
    \item Sample start positions $t_1, t_2$ in primary and secondary
    \item Extract regions: $r_1 \gets x_1[t_1:t_1+L]$, $r_2 \gets x_2[t_2:t_2+L]$
    \item Compute energies $E_1, E_2$
    \item Sample energy ratio: $\rho \sim \text{Uniform}(-5, 5)$ dB
    \item Compute scale: $\beta \gets \sqrt{E_1 \cdot 10^{\rho/10} / E_2}$
    \item Mix: $x_1[t_1:t_1+L] \gets r_1 + \beta \cdot r_2$
\end{enumerate}

Mixing probability is 0.25 with max 50\% overlap.

\subsection{Augmentor Buffer}

The augmentor maintains a buffer of 64 recent utterances for noise/mixing sources, updated after each batch and stored on CPU.

\begin{table}[t]
  \caption{SER accuracy (\%) per fold for $\lambda_{\text{end}}=0$ ablation.}
  \label{tab:ablation_ser_folds}
  \begin{center}
    \begin{tiny}
      \begin{sc}
        \begin{tabular}{lcccccr}
          \toprule
          $\lambda_{\text{end}}$ & F1 & F2 & F3 & F4 & F5 & Avg \\
          \midrule
          0.00 & 63.48 & 63.41 & 63.70 & 63.22 & 65.15 & 63.79 \\
          0.01 & \textbf{66.07} & \textbf{67.85} & \textbf{66.67} & \textbf{68.56} & \textbf{67.37} & \textbf{67.30} \\
          \bottomrule
        \end{tabular}
      \end{sc}
    \end{tiny}
  \end{center}
\end{table}

\section{Visualization Details}
\label{app:visualization}

\paragraph{UMAP Projection.}
Table~\ref{tab:umap_params} lists the hyperparameters used for UMAP projections in Figure~\ref{fig:umap}.

\begin{table}[h]
  \caption{UMAP hyperparameters.}
  \label{tab:umap_params}
  \begin{center}
    \begin{small}
      \begin{sc}
        \begin{tabular}{lc}
          \toprule
          Parameter & Value \\
          \midrule
          n\_neighbors & 15 \\
          min\_dist & 0.1 \\
          metric & Euclidean \\
          n\_components & 2 \\
          random\_state & 42 \\
          n\_samples & 10,000 \\
          \bottomrule
        \end{tabular}
      \end{sc}
    \end{small}
  \end{center}
\end{table}

\end{document}